\documentclass[lettersize,journal]{IEEEtran}
\usepackage{amsmath,amsfonts}
\usepackage{algorithmic}
\usepackage{algorithm}
\usepackage{array}
\usepackage[caption=false,font=normalsize,labelfont=sf,textfont=sf]{subfig}
\usepackage{textcomp}
\usepackage{stfloats}
\usepackage{multirow}
\usepackage{url}
\usepackage{verbatim}
\usepackage{graphicx}
\usepackage{hyperref}
\usepackage{tabularx}
\usepackage[
backend=biber,
sorting=ynt
]{biblatex}

\begin{document}

\title{Dual Path Learning - learning from noise and context for medical image denoising}

\author{Jitindra Fartiyal, Pedro Freire, Yasmeen Whayeb, James S. Wolffsohn, Sergei K. Turitsyn, Sergei G. Sokolovski}



\maketitle

\begin{abstract}
Medical imaging plays a critical role in modern healthcare, enabling clinicians to accurately diagnose diseases and develop effective treatment plans. However, noise, often introduced by imaging devices, can degrade image quality, leading to misinterpretation and compromised clinical outcomes. Existing denoising approaches typically rely either on noise characteristics or on contextual information from the image. Moreover, they are commonly developed and evaluated for a single imaging modality and noise type. Motivated by CNCL \cite{geng_content-noise_2022}, which integrates both noise and context, this study introduces a Dual-Pathway Learning (DPL) model architecture that effectively denoises medical images by leveraging both sources of information and fusing them to generate the final output. DPL is evaluated across multiple imaging modalities and various types of noise, demonstrating its robustness and generalizability. DPL improves PSNR by 3.35\% compared to the baseline UNet when evaluated on Gaussian noise and trained across all modalities. The code is available at 10.5281/zenodo.15836053."
\end{abstract}


\footnote{Jitindra Fartiyal  (240304225@aston.ac.uk), Pedro Friere (\texttt{p.freiredecarvalhosourza@aston.ac.uk}), Sergei K. Turitsyn (\texttt{s.k.turitsyn@aston.ac.uk}), and Sergei G. Sokolovski (\texttt{s.sokolovsky@aston.ac.uk}) is with Aston Institute of Photonics Technology (AIPT) at Aston University, Birmingham, UK. Yasmeen Whayeb (\texttt{149007690@aston.ac.uk}) is with HLS\_HLSC at Aston University, Birmingham, UK. James S. Wolffsohn (\texttt{J.S.W.Wolffsohn@aston.ac.uk}) is with School Of Optometry at Aston University, Birmingham, UK.}

\section{Introduction}
Medical imaging plays an important role in the diagnosis, monitoring, and treatment of various diseases. It enables the visualization of the body's anatomy, and physiological functions, helping healthcare professionals to make accurate clinical decisions. Additionally, medical imaging is essential for tracking the progression of conditions like brain tumors, allowing timely and appropriate interventions. A major challenge in medical imaging is the presence of noise in the image, which can compromise image clarity which is an essential factor for accurate and reliable clinical diagnosis. In modalities like CT scans, where full-dose scans poses risks due to higher radiation exposure, low-dose CT scans offer a safer alternative. However, these low-dose scans are naturally more prone to noise. Additionally, image quality significantly affects the performance of artificial intelligence algorithms and diagnostic tools. Therefore, obtaining high-quality images is critical for effective analysis and reliable outcomes.

Noise in medical images can arise from several sources, including human errors—such as patient movement and variations in operator protocols—and non-human factors, such as the intrinsic noise associated with specific medical imaging modalities. While human error can often be reduced through intensive training and standardized procedures, this study focuses on the inherent noise present across different imaging techniques. Various imaging modalities exhibit distinct types of noise. For example, Additive White Gaussian noise is commonly found in CT scans, while OCT images are often affected by Speckle noise \cite{nazir_recent_2024}.

Medical images often contain a combination of diverse noise types. The most common types include gaussian, Speckle, Poisson, and Salt \& pepper noise. Gaussian noise follows a normal probability distribution and is commonly encountered in CT and MRI scans \cite{nazir_recent_2024}. It introduces random fluctuations in pixel intensities, with the magnitude of these fluctuations determined by the variance of the associated distribution. Speckle noise, frequently observed in ultrasound and OCT scans \cite{nazir_recent_2024}, is characterized by its multiplicative nature and the granular texture it imparts to images. This type of noise arises from the constructive and destructive interference of light waves. Poisson noise, also called quantum noise, is signal-dependent and stems from the fundamental characteristics of electromagnetic radiation \cite{kaur_complete_2023}. It is most commonly observed in imaging modalities such as X-ray and PET scans \cite{nazir_recent_2024}. Salt-and-pepper noise, also known as impulsive noise, randomly introduces bright and dark pixels across the image \cite{kaur_complete_2023}. It is frequently encountered in imaging techniques such as X-ray, MRI, and CT scans \cite{nazir_recent_2024}.

Existing research has already established the efficacy of deep learning-based denoising over traditional denoising algorithms. Deep learning-based denoising models are typically built on Convolutional Neural Networks (CNNs), Generative Adversarial Networks (GANs), and diffusion models. All these architectures have shown significant advancements, particularly GANs; however, they often suffer from complex training procedures and intensive parameter tuning. In contrast, CNN-based models offer competitive performance with simpler training and reduced complexity. Building on this, CNN architectures can be further be optimized to enhance denoising performance. Inspired by the approaches proposed in \cite{sharif_two-stage_2024} and \cite{geng_content-noise_2022}, we propose a dual-path pipeline that separately processes noise and contextual information within the image, which are then fused to reconstruct the denoised output. The architecture diagram is shown in Figure \ref{fig:dpl_architecture}.

The main contributions of this study are as follows:
\begin{itemize}
    \item To the best of our knowledge, this study performs large-scale, diverse training and validation on datasets containing three different types of noise across four distinct medical imaging modalities.
    \item This research proposes an extension of the CNCL algorithm \cite{geng_content-noise_2022}, which incorporates a GAN-based loss function. The proposed DPL model leverages a convolutional neural network to fuse information from both noise and contextual cues—not to generate an image directly, but to integrate informative features. This focus on information fusion differentiates our approach from conventional image generation models. Additionally, our generalized architecture enables experimentation with various encoder-decoder blocks, such as autoencoders, providing flexibility for future improvements.
\end{itemize}

Section \pageref{related-works} discusses the related work, and Section \pageref{methodology} details the data preparation and model. Section \pageref{results} presents the experimental results, which are then followed by Section \pageref{discussions}, which discusses insights from this research. The study is concluded in Section \pageref{conclusion}.

\section{Related works} \label{related-works}
Medical imaging is undeniably vital in healthcare, yet noise remains an unavoidable aspect of these images. Therefore, producing noise-free images is essential for accurate diagnosis and effective treatment. While traditional denoising methods exist, they often suffer from high sensitivity to image variations, leading to over-denoising and the loss of critical structural details. Although some deep learning-based denoising approaches have been proposed, they are typically designed for specific image modalities and noise types, limiting their generalizability to other medical imaging modalities. This lack of versatility hinders their practical application in real-world clinical settings.

Classical denoising techniques primarily utilize neighborhood-based and transform-domain approaches. Table \ref{tab:classical-denoising-algorithms} summarizes these algorithms along with their respective advantages and disadvantages.

\begin{table*}[ht!]
    \centering
    \caption{Summary of Classical Denoising Techniques: Underlying Principles, Strengths, and Weaknesses}
    \begin{tabular}{m{6em}  m{35em}  m{10em}  m{10em} }
        \hline
        Algorithm & Definition & Advantages & Disadvantages \\
        \hline
        Mean Filter & Replaces each pixel as a mean of a N x N neighborhood. & Simple \& fast. & Over smoothening. \\
        Median Filter & Replaces each pixel as a median of a N x N neighborhood. & Better than mean filter near edges. & Over smoothening. \\
        Gaussian Filter & Applies a Gaussian kernel in a N x N neighborhood. & Effective in reducing gaussian noise. & Bad in preserving edges and textures. \\
        Bilateral Filter & Similar to Gaussian filter but combines spatial and intensity weighting. & Preserves edges better. & Parametric sensitive. \\
        Wavelet Denoising \cite{ruikar_image_2010} & Transform image into wavelet coefficients, apply thresholds and reconstruct. & Good in preserving fine details. & Threshold sensitive. \\
        Non-Local Means \cite{buades_non-local_2011} & Extends beyond local neighborhoods by identifying similar patches across the image and average the intensity. & Good in preserving fine details. & Computationally intensive. \\
        Block-Matching 3D (BM3D)  \cite{dabov_image_nodate} & Grouping similar blocks into 3D volumes and performing collaborative filtering in the transform domain, followed by inverse transformation and optional Wiener filtering for refinement. & State of the art in classical algorithms. & Computationally intensive and complex. \\
        \hline
    \end{tabular}
    \label{tab:classical-denoising-algorithms}
\end{table*}

Recent advances in deep learning have significantly improved image denoising, particularly in medical imaging. Various architectures including Convolutional Neural Networks (CNNs), Generative Adversarial Networks (GANs), Transformers, and Diffusion models have been proposed to address challenges related to noise, artifact preservation, and feature retention. Chen \cite{chen_low-dose_2017} introduced RED-CNN, leveraging a symmetric residual autoencoder with shortcut connections to denoise low-dose CT images effectively while preserving structural details. Sharif \cite{sharif_learning_2020} proposed the Dynamic Residual Attention Network (DRAN), which employs dynamic convolutions and spatial attention to enhance feature learning and mitigate over-smoothing. In a later work, Sharif \cite{sharif_two-stage_2024} developed a two-stage model incorporating residual dense blocks and a noise attention mechanism for better pixel-level noise estimation. Dong \cite{dong_feature-guided_2021} presented FDCNN, a feature guided CNN for ultrasound denoising using guided backpropagation and dilated convolutions. Yang \cite{yang_low-dose_2023} proposed a dual-network approach-Feature Refinement Network (FRN) and Dynamic Perception Network (DPN), utilizing residual dense blocks and dynamic convolution for improved multiscale feature processing and denoising performance.

Genf \cite{geng_content-noise_2022} introduced a GAN-based model with separate content and noise predictors fused and validated using a PatchGAN discriminator, outperforming traditional L1/L2-based methods. Huang \cite{huang_du-gan_2022} extended this with a dual-domain GAN framework using RED-CNN as a generator and U-Net discriminators in both image and gradient domains to enhance detail and edge preservation. Mishra \cite{mishra_ultrasound_2018} presented a GAN architecture for ultrasound despeckling using a deep residual generator and used adversarial and structural losses to maintain image fidelity.

To overcome the limitations of CNNs and GANs, CSformer \cite{yin_csformer_2022} integrates cross-scale fusion and Swin Transformer blocks for enhanced global-local context modeling. TECDNet \cite{zhao_hybrid_2022}, which combines a Swin Transformer encoder with a convolutional decoder and radial basis function attention, further advances denoising capabilities. Additionally, a despeckling pipeline \cite{hu_retinal_2020} employing a pseudo-modality concept and a pseudomultimodal fusion network (PMFN) has demonstrated effectiveness by incorporating a robust edge map as a structural prior.

\section{Methodology} \label{methodology}
In this section, we detail the evaluation metrics employed, data preparation stages, model architecture, learning objectives, and training parameters.
\subsection{Evaluation Metrics}
\subsubsection{Mean Squared Error}
Mean Squared Error (MSE) measures the average squared difference between the noisy image and its noise-free counterpart. It evaluates only pixel intensity differences, without accounting for local structures or important regions within the image. Lower MSE values indicate more effective denoising, while higher values suggest poorer performance. Equation \ref{eq:mse} defines the MSE as:
\begin{equation}
    MSE = \frac{1}{n}\sum^n_{i=1}(Y_i - Y'_i)^2
    \label{eq:mse}
\end{equation}
where $Y'_i$ and $Y_i$ is the noisy and noise-free $i^{th}$ image and $n$ is the total number of images.

\subsubsection{Peak Signal to Noise Ratio}
Peak Signal-to-Noise Ratio (PSNR) represents the ratio between the maximum possible signal power (typically 255 for images) and the Mean Squared Error (MSE) \cite{kaur_complete_2023}, \cite{hore_image_2010}. It is commonly used to assess the level of noise or distortion introduced during the denoising process. A higher PSNR value signifies better denoising performance. Equation \ref{eq:psnr} defines PSNR.
\begin{equation}
    PSNR = 10log_{10}\frac{maximum^{Intensity}}{MSE}
    \label{eq:psnr}
\end{equation}
where, $maximum^{Intensity}$ is the maximum possible intensity of the image, and MSE is the mean squared error between noisy and noise-free image.

\subsubsection{Structural Similarity}
The Structural Similarity Index (SSIM) is a perceptual metric used to evaluate the similarity between noisy and noise-free images by considering contrast, luminance, and structural information \cite{kaur_complete_2023}, \cite{wang_image_2004}. Its values range from 0 to 1, where 1 denotes perfect similarity and 0 indicates no similarity. Unlike conventional metrics, SSIM aligns more closely with human visual perception. Equations \ref{eq:ssim-l}, \ref{eq:ssim-c}, \ref{eq:ssim-s}, and \ref{eq:ssim} define the components and overall calculation of SSIM:
\begin{equation}
    l(x,y) = \frac{2\mu_x\mu_y + c_1}{\mu_x^2 + \mu_y^2 + c_1}
    \label{eq:ssim-l}
\end{equation}
\begin{equation}
    c(x,y) = \frac{2\sigma_x\sigma_y + c_2}{\sigma_x^2 + \sigma_y^2 + c_2}
    \label{eq:ssim-c}
\end{equation}
\begin{equation}
    s(x,y) = \frac{\sigma_{xy} + c_3}{\sigma_x\sigma_y + c_3}
    \label{eq:ssim-s}
\end{equation}
\begin{equation}
    SSIM = l(x,y)^\alpha * c(x,y)^\beta * s(x,y)^\gamma
    \label{eq:ssim}
\end{equation}
Here, $\alpha$, $\beta$, and $\gamma$ are the weights of the score use to compute SSIM. $l(x,y)$, $c(x,y)$, and $s(x,y)$ is the luminance, contrast, and structure between two images: $x$, and $y$ which is a pair of noisy and noise-free images.

\subsection{Data Preparation}
To evaluate the DPL, five medical imaging datasets were used. Four of these represent separate datasets across different imaging modalities: CT, MRI, OCT, and Fundus. In addition, these datasets were combined into a single dataset to evaluate the generalizability of the model across different imaging modalities. To ensure consistent algorithmic performance, the dataset comprises medical images from a diverse patient population with a range of health conditions. Furthermore, each imaging modality is represented by a substantial number of samples (approximately 500 images per modality except for OCT), promoting generalizability within individual modalities. This design choice also helps to mitigate potential biases arising from class imbalance across different image modalities. While most existing denoising research focuses on a single medical imaging modality, our dataset addresses this limitation by incorporating four distinct types of medical images. To address the lack of paired noisy and noise-free images, a synthetic noise generation process was employed. Additionally, most existing denoising models are limited to handling a single noise type, restricting their generalizability. To overcome this limitation, the proposed dataset incorporates three distinct types of noise: Gaussian, additive white Gaussian noise (AWGN), and speckle noise.

\subsubsection{Data Collection}
The datasets originate from both publicly available repositories and private sources. A detailed summary, including dataset size, anatomical focus, disease categories, and data source, is provided in Table \ref{tab:data_collection_table}.

\begin{table*}[ht!]
    \centering
    \caption{Comprehensive Summary of the dataset used}
    \begin{tabular}{m{7em} m{5em}  m{5em}  m{7em} m{5em}  m{15em}  m{5em}}
        \hline
        Image Modality & Dataset Source & Size & Anatomical Structure & \#Patients & Disease Focus & Source Type  \\
        \hline
        CT & LDCT 2016 \cite{mccollough_low_2020} & 650 & Chest & 5 & Two healthy \& three with pulmonary nodule. & Public \\
        MRI & BraTS Africa 2021 \cite{adewole_brain_2023} & 583 & Brain & 146 & Glioblastoma and other Central Nervous System neoplasms. & Public \\
        FUNDUS & ORIGA \cite{zhang_origa-light_2010} & 650 & Eye & 650 & Glaucoma. & Public \\
        OCT & Aston & 12 & Eye & Not available & Healthy individuals & Private \\ 
        OCT & PKU 37 \cite{geng_triplet_2022} & 33 & Eye & 33 & Healthy individuals & Public \\
        OCT & SDOCT \cite{fang_sparsity_2012} & 17 & Eye & 17 & Healthy and pathological individuals & Public \\ 
        \hline
    \end{tabular}
    \label{tab:data_collection_table}
\end{table*}

\subsubsection{Data Preproccesing}
Each dataset was manually inspected for corrupted images, which were then removed. Subsequently, all medical images were saved as NumPy files with a size of 256×256 pixels to facilitate faster read and write operations from memory.

Chest CT datasets require data sampling because they consist of volumetric scans. We utilize 130 scans from five patients to construct the datasets. To ensure complete coverage of each patient's chest CT volume, we sample 130 scans per patient from the full set of images.

The MRI scan dataset comprises T1-contrast (T1c), T1-native (T1n), T2, and T2-FLAIR types of images. All these MRI scan types are employed in the dataset creation process.

Typically, the dataset is divided into training and validation sets in an 80:20 ratio. For the CT scans, 30 scans from each of the five patients were randomly selected as the validation set, while the remaining scans were utilized as the training set. For MRI, one scan randomly selected from T1n, T1c, T2, and T2-FLAIR scans from each of the 146 patients was selected for validation, while the remaining scans were used for training. For OCT, 150 patients’ scans were randomly selected for the validation set, with the remaining patients’ scans used for training. An 80:20 ratio is applied to split the dataset across each OCT data source.

\begin{table}[ht!]
    \centering
    \caption{Details of various types of Noise added to the dataset}
    \begin{tabular}{m{3.5em} m{4em} | m{4em} m{4em} | m{4em} m{3.5em}}
        \hline
        Parameters & Gaussian & Parameters & AWGN & Parameters & Speckle \\
        \hline
        Mean & 0 & Covar:loc & 0.01 & Mean & 0.1 \\
        Var & 0.005 & Covar:scale & 0.0001 & Var & 0.01 \\
        \hline
    \end{tabular}
    \label{tab:noise-simulation-details}
\end{table}

\begin{figure}[ht!]
    \centering
    \includegraphics[width=1.0\linewidth]{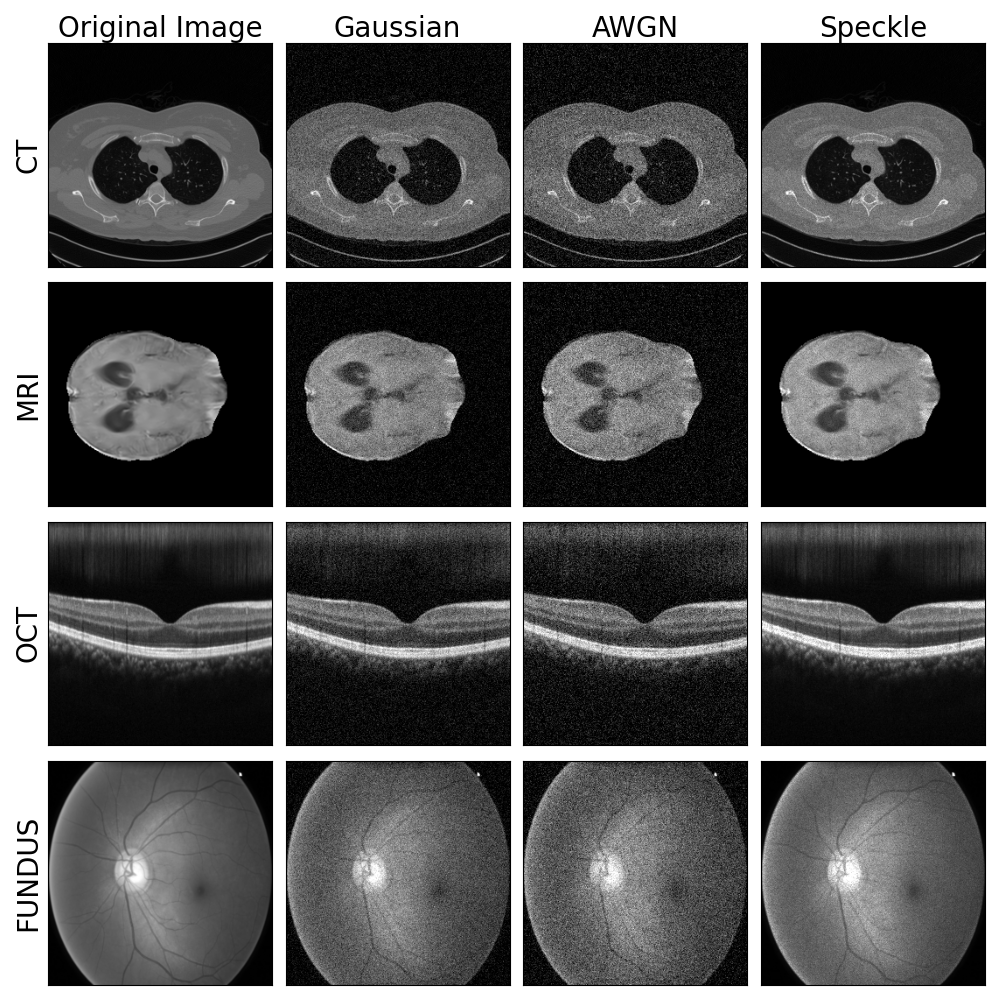}
    \caption{Noisy CT, MRI, OCT, and Fundus images after adding Gaussian, AWGN, and Speckle noise}
    \label{fig:noise-simulation-viz}
\end{figure}

\subsubsection{Noise Simulation}
Noise is added at multiple levels across all sources of noise. Three types of noise addition — Gaussian, additive white Gaussian noise (AWGN), and speckle noise are implemented using the scikit-learn library across all medical imaging modalities. Figure \ref{fig:noise-simulation-viz} illustrates different types of noise applied to various imaging modalities. Table \ref{tab:noise-simulation-details} provides further details about the noise.

\subsection{Dual-Path Learning}
\begin{figure*}[ht!]
    \centering
    \includegraphics[scale=0.32]{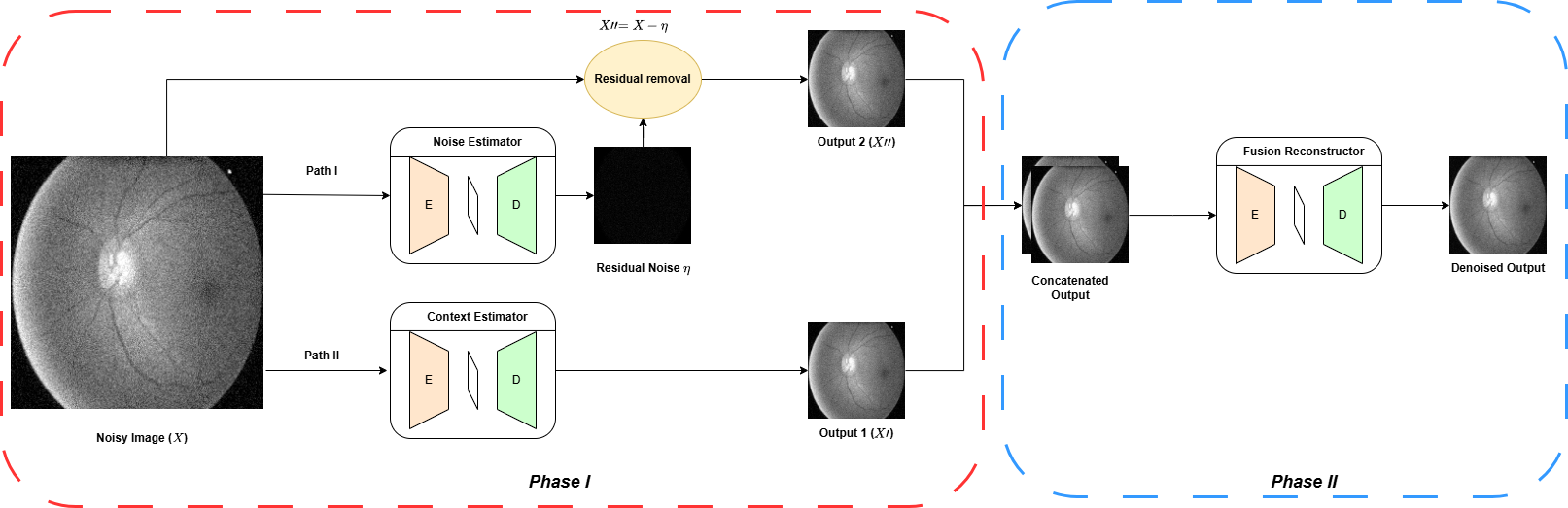}
    \caption{The proposed two-phase architecture starts with a dual-path design, where each path learns noise and contextual information in the image. In the second phase, the intermediate outputs from both paths are concatenated and fused to produce the final denoised image.}
    \label{fig:dpl_architecture}
\end{figure*}

This study introduces a Dual-Path Learning framework comprising two phases. The overall architecture is illustrated in Figure \ref{fig:dpl_architecture}.

In the first phase, the model comprises two parallel paths: a noise estimator and a context estimator. The noise estimator estimates the noise, which is then used to derive an intermediate denoised output \(X''\), as described by Equation \ref{eq:noise-to-context_equation}. The noise estimator is implemented as a standard autoencoder network, specifically a vanilla U-Net in this study.
\begin{equation}
    X'' = X - \eta
    \label{eq:noise-to-context_equation}
\end{equation}
Simultaneously, the context estimator recovers the underlying clean context \(X'\) directly from the noisy input \(X\) using a vanilla U-Net model similar to the noise estimator. In the second phase, the outputs from both estimators, \(X'\) and \(X''\) are concatenated and passed to a fusion reconstructor. This reconstructor, also implemented as a vanilla U-Net autoencoder, fuses the complementary information from both paths to generate the final denoised image \(Y'\). As mentioned earlier, all autoencoder blocks employ the U-Net architecture by Ronneberger et al. \cite{ronneberger_u-net_2015}, with the only modification being that the last layer outputs a single channel instead of two, as in the original design.

\subsection{Learning Objectives}
The proposed Dual-Path Learning network uses three distinct loss functions. The learning objective \(L_o\) is defined in equation \ref{eq:loss_function}
\begin{equation}
    L_o = argmin_w (L_n + L_c + L_f)
    \label{eq:loss_function}
\end{equation}
where \(L_n\), \(L_c\), and \(L_f\) are the Minimum Squared Error(MSE) loss for noise, context and fusion reconstructor network. Equation \ref{eq:noise_loss}, \ref{eq:context_loss}, and \ref{eq:fusion_recon_loss} defines the respective loss equations
\begin{equation}
    L_n = \frac{1}{b}\sum^b (Y - X'')^2
    \label{eq:noise_loss}
\end{equation}

\begin{equation}
    L_c = \frac{1}{b}\sum^b (Y - X')^2
    \label{eq:context_loss}
\end{equation}

\begin{equation}
    L_f = \frac{1}{b}\sum^b (Y - Y')^2
    \label{eq:fusion_recon_loss}
\end{equation}
where \(b\) is the batch size and \(Y\) is the ground truth.

\subsection{Training Details}
The Dual-Path Learning model is implemented using the PyTorch framework. As described earlier, a U-Net architecture is employed at all stages of the model. Each U-Net is optimized using the Adam optimizer with a learning rate of \(1*e^{-4}\). The DPL model is trained for 10,000 epochs with a batch size of 32, using input images of size \(256*256\). Training was conducted on NVIDIA RTX A6000 GPUs with 48 GB of memory and took approximately 46 hours to complete when trained solely on CT images. When trained on all medical imaging modalities combined, training time increased to approximately 133 hours. For comparison, baseline models including a standalone U-Net and the RED-CNN architecture were trained under the same number of epochs (10,000) and learning rate ( \(1*e^{-4}\) but with a batch size of 64.
 
\section{Results} \label{results}
\begin{figure} [ht!]
    \centering
    \includegraphics[width=1.0\linewidth]{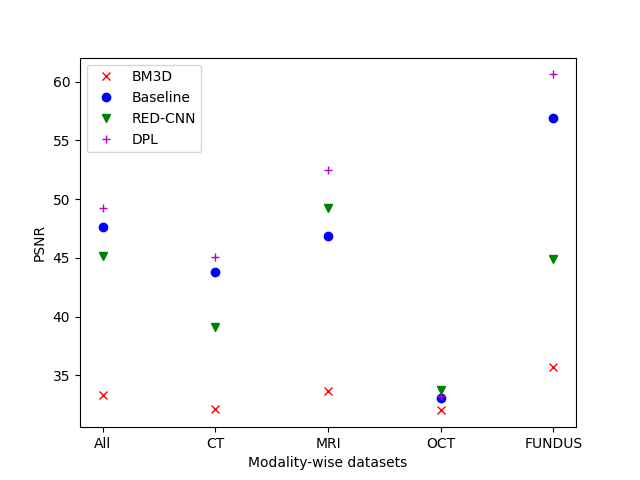}
    \caption{PSNR comparison of BM3D, baseline U-Net, RED-CNN, and DPL under Gaussian noise across five medical imaging datasets.}
    \label{fig:comparative-psnr-results}
\end{figure}

\begin{figure}[ht!]
    \centering
    \includegraphics[width=1.0\linewidth]{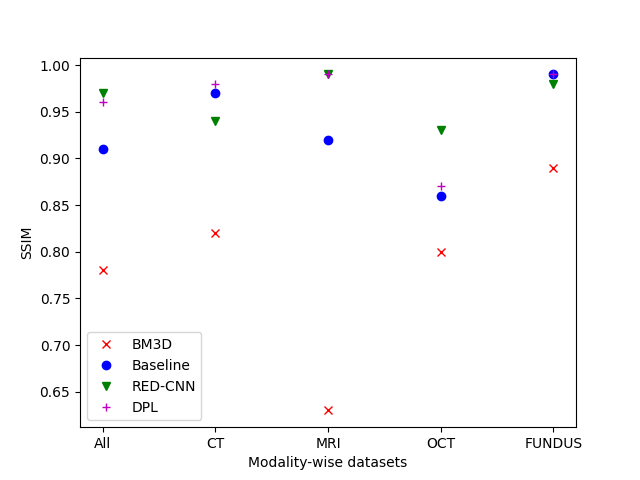}
    \caption{SSIM comparison of BM3D, baseline U-Net, RED-CNN, and DPL under Gaussian noise across five medical imaging datasets.}
    \label{fig:ssim-results}
\end{figure}

This study evaluates the performance of the proposed Dual-Path Learning model across five distinct medical imaging datasets. The comparison includes several classical denoising algorithms — Median, Gaussian, Bilateral Filtering, Wavelet Filtering, Wiener Filtering, Non-Local Means (NLM), and BM3D as well as the state-of-the-art deep learning method RED-CNN \cite{chen_low-dose_2017}. Table \ref{tab:psnr-results} presents the comparative results using PSNR as the evaluation metric under Gaussian, additive white Gaussian noise (AWGN), and speckle noise conditions. Figure \ref{fig:comparative-psnr-results} illustrates the comparative PSNR performance of BM3D, the baseline U-Net, RED-CNN, and DPL under Gaussian noise. DPL improves PSNR by 3.35\% over the baseline, 9\% over RED-CNN, and 47.6\% over BM3D under Gaussian noise. Among all imaging modalities, the greatest improvement is observed for FUNDUS images, with a 69.8\% gain over BM3D, while OCT shows the least improvement under the same noise condition. Under AWGN and speckle noise, DPL improves PSNR by 9.6\% and 22.34\% over BM3D, respectively, less pronounced than its gains under Gaussian noise. For these noise types, DPL does not surpass RED-CNN but still outperforms the baseline and most classical methods across the majority of modalities. These findings demonstrate that DPL performs exceptionally well against Gaussian noise and remains competitive under other noise conditions, exhibiting strong generalizability across multiple modalities. In addition to PSNR, the Structural Similarity Index (SSIM) was used for further evaluation. Figure \ref{fig:ssim-results} visualizes the SSIM scores across all five datasets. While no decisive performance is observed for DPL, it still consistently outperforms BM3D. Table \ref{tab:ssim-results} provides comprehensive SSIM-based results across all noise types.

\begin{table*}[ht!]
    \centering
    \caption{PSNR comparison of denoising algorithms across imaging modalities under Gaussian, AWGN, and speckle noise.}
    \begin{tabular}{m{6em}   m{2em} m{2em} m{2em} m{2em} m{3.5em} |  m{2em} m{2em} m{2em} m{2em} m{3.5em} | m{2em} m{2em} m{2em} m{2em} m{3.5em}}
        \hline
        \multicolumn{1}{c}{} & \multicolumn{5}{ c }{Gaussian} & \multicolumn{5}{ c }{AWGN} & \multicolumn{5}{ c }{Speckle} \\
        \hline
        Algorithm & All & CT & MRI & OCT & FUNDUS & All & CT & MRI & OCT & FUNDUS & All & CT & MRI & OCT & FUNDUS \\
        \hline
        \hline
        Median filter & 28.79 & 26.22 & 30.46 & 29.03 & 29.46 & 27.49 & 25.19 & 28.92 & 27.57 & 28.27 & 28.81 & 28.18 & 29.62 & 30.54 & 26.89\\
        Gaussian filter & 28.18 & 25.92 & 28.36 & 29.53 & 28.91 & 26.51 & 24.53 & 26.24 & 28.04 & 27.24 & 28.01 & 26.51 & 29.09 & 30.37 & 26.05 \\
        Bilateral filter & 28.37 & 26.57 & 28.42 & 29.35 & 29.13 & 26.51 & 24.83 & 26.22 & 27.70 & 27.31 & 28.64 & 27.67 & 29.51 & 30.97 & 26.41 \\
        Wavelet (B) & 27.21 & 25.81 & 26.79 & 29.04 & 27.18 & 25.46 & 23.79 & 24.83 & 27.20 & 26.01 & 27.50 & 28.74 & 24.85 & 30.19 & 26.24 \\
        Wavelet (V) & 21.00 & 18.40 & 21.91 & 23.78 & 19.90 & 20.08 & 17.08 & 20.90 & 22.93 & 19.43 & 24.25 & 24.94 & 22.13 & 28.65 & 21.30 \\
        Wiener filter & 30.57 & 29.50 & 29.73 & 30.99 & 32.06 & 27.80 & 26.88 & 26.72 & 28.69 & 28.92 & 28.71 & 28.63 & 29.70 & 29.94 & 26.58 \\
        NLM & 30.57 & 30.67 & 29.12 & 29.92 & 32.56 & 28.20 & 28.17 & 26.55 & 28.02 & 30.07 & 28.71 & 28.54 & 29.99 & 29.53 & 26.77 \\
        BM3D & 33.35 & 32.10 & 33.63 & 32.01 & 35.69 & 30.20 & 29.57 & 29.50 & 30.65 & 31.10 & 30.02 & 30.10 & 30.42 & 30.77 & 28.81 \\
        \hline
        Baseline & 47.65 & 43.76 & 46.87 & 33.10 & 56.94 & 33.17 & 32.95 & 32.35 & 32.17 & 35.10 & 36.61 & 37.30 & 36.05 & 34.92 & 38.85 \\
        REDCNN & 45.18 & 39.15 & 49.28 & \textbf{37.77} & 44.93 & \textbf{35.99} & \textbf{34.29} & \textbf{36.41} & \textbf{33.96} & \textbf{38.12} & \textbf{40.89} & \textbf{39.88} & \textbf{41.39} & \textbf{38.12} & \textbf{41.82} \\
        DPL & \textbf{49.25} & \textbf{45.03} & \textbf{52.51} & 33.23 & \textbf{60.62} & 33.87 & 33.64 & 33.49 & 32.11 & 35.57 & 37.12 & 37.48 & 35.93 & 35.76 & 38.96 \\
        \hline
    \end{tabular}
    \label{tab:psnr-results}
\end{table*}

\begin{table*}[ht!]
    \centering
    \caption{SSIM comparison of denoising algorithms across imaging modalities under Gaussian, AWGN, and speckle noise.}
    \begin{tabular}{m{6em}   m{2em} m{2em} m{2em} m{2em} m{3.5em} |  m{2em} m{2em} m{2em} m{2em} m{3.5em} | m{2em} m{2em} m{2em} m{2em} m{3.5em}}
        \hline
        \multicolumn{1}{c}{} & \multicolumn{5}{ c }{Gaussian} & \multicolumn{5}{ c }{AWGN} & \multicolumn{5}{ c }{Speckle} \\
        \hline
        Algorithm & All & CT & MRI & OCT & FUNDUS & All & CT & MRI & OCT & FUNDUS & All & CT & MRI & OCT & FUNDUS \\
        \hline
        \hline
        Median filter & 0.64 & 0.62 & 0.59 & 0.64 & 0.7 & 0.56 & 0.52 & 0.51 & 0.62 & 0.61 & 0.89 & 0.85 & 0.93 & 0.90 & 0.87 \\
        Gaussian filter & 0.59 & 0.62 & 0.36 & 0.69 & 0.68 & 0.50 & 0.51 & 0.30 & 0.62 & 0.57 & 0.90 & 0.87 & 0.94 & 0.92 & 0.85 \\
        Bilateral filter & 0.56 & 0.59 & 0.35 & 0.67 & 0.65 & 0.47 & 0.47 & 0.29 & 0.59 & 0.54 & 0.89 & 0.87 & 0.94 & 0.92 & 0.84 \\
        Wavelet (B) & 0.58 & 0.58 & 0.30 & 0.70 & 0.73 & 0.53 & 0.51 & 0.26 & 0.66 & 0.68 & 0.73 & 0.85 & 0.41 & 0.91 & 0.76 \\
        Wavelet (V) & 0.52 & 0.49 & 0.27 & 0.60 & 0.70 & 0.50 & 0.45 & 0.25 & 0.61 & 0.68 & 0.66 & 0.75 & 0.32 & 0.80 & 0.77 \\
        Wiener filter & 0.66 & 0.73 & 0.37 & 0.76 & 0.78 & 0.56 & 0.62 & 0.28 & 0.68 & 0.65 & 0.91 & 0.88 & 0.95 & 0.91 & 0.91 \\
        NLM & 0.62 & 0.73 & 0.34 & 0.72 & 0.72 & 0.53 & 0.62 & 0.26 & 0.63 & 0.60 & 0.86 & 0.82 & 0.94 & 0.88 & 0.81 \\
        BM3D & 0.78 & 0.82 & 0.63 & 0.80 & 0.89 & 0.63 & 0.72 & 0.37 & 0.75 & 0.71 & 0.86 & 0.88 & 0.78 & 0.86 & 0.93 \\
        \hline
        Baseline & 0.91 & 0.97 & 0.92 & 0.86 & 0.99 & 0.80 & 0.88 & 0.60 & 0.86 & 0.92 & 0.88 & \textbf{0.96} & 0.91 & 0.90 & 0.95 \\
        REDCNN & \textbf{0.97} & 0.94 & \textbf{0.99} & \textbf{0.93} & 0.98 & \textbf{0.91} & 0.88 & \textbf{0.95} & \textbf{0.89} & \textbf{0.93} & \textbf{0.97} & \textbf{0.96} & \textbf{0.98} & \textbf{0.96} & \textbf{0.96} \\
        DPL & 0.96 & \textbf{0.98} & \textbf{0.99} & 0.87 & \textbf{0.99} & 0.84 & \textbf{0.90} & 0.88 & 0.85 & 0.91 & 0.95 & \textbf{0.96} & 0.88 & 0.93 & 0.95 \\
        \hline
    \end{tabular}
    \label{tab:ssim-results}
\end{table*}

\begin{figure*}
    \centering
    \includegraphics[width=1.0\linewidth]{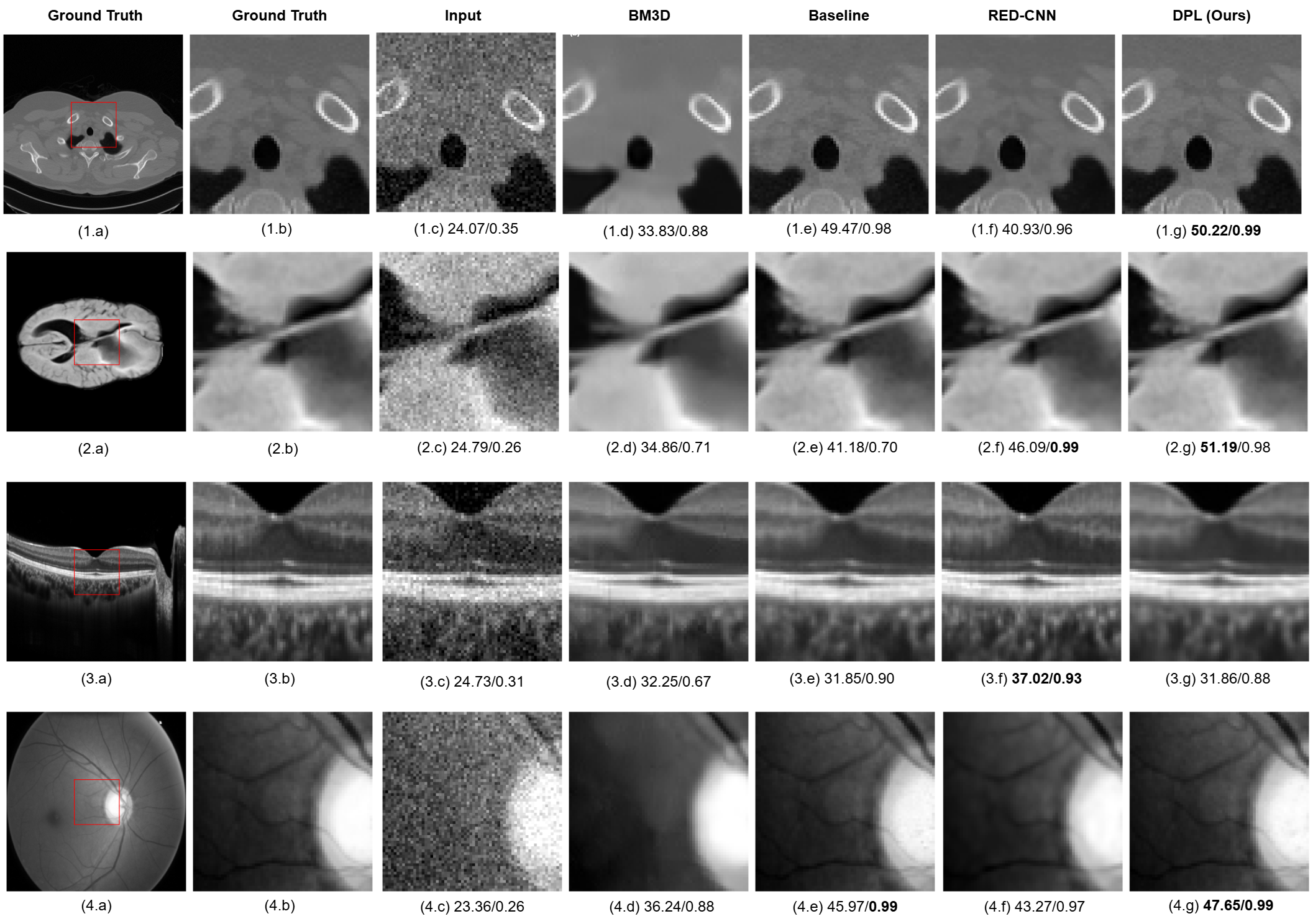}
    \caption{Visualization of CT, MRI, OCT, FUNDUS sample results before and after denoising on against Gaussian Noise. Top row denoted as (1.x) presents CT, (2.x) presents MRI, (3.x) presents OCT, and (4.x) presents FUNDUS. (x.a) and (x.b) presents the full and zoomed part of the ground truth. (x.c), (x.d), (x.e), (x.f) presents results of BM3D, baseline, RED-CNN and DPL algorithms.}
    \label{fig:full-vis}
\end{figure*}

Figures \ref{fig:full-vis} present both full and zoomed-in visualizations of the results produced by various algorithms on the CT, MRI, OCT, and Dundus datasets when trained under Gaussian noise. All algorithms reduce noise to some extent. However, BM3D tends to over-smooth the images, leading to the loss of fine structural details. The baseline model retains structural information but introduces noticeable streak like artifacts. Both RED-CNN and DPL perform effectively; however, DPL offers improved contrast, particularly in zoomed-in regions and avoids excessive smoothing. It also demonstrates superior preservation of dark regions within the images. Overall, DPL yields the most visually compelling results among the compared methods.

\section{Discussions} \label{discussions}
\begin{figure}[h!]
    \centering
    \includegraphics[width=0.5\textwidth]{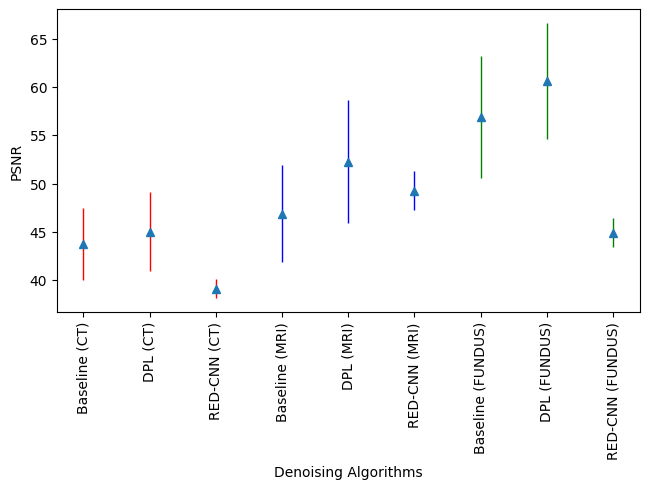}
    \caption{Comparison of mean and standard deviation in PSNR results for Baseline, RED-CNN, and DPL models across CT, MRI, OCT, and fundus datasets under Gaussian noise.}
    \label{fig:mean-var-res}
\end{figure}

\begin{figure}[h!]
    \centering
    \includegraphics[width=0.5\textwidth]{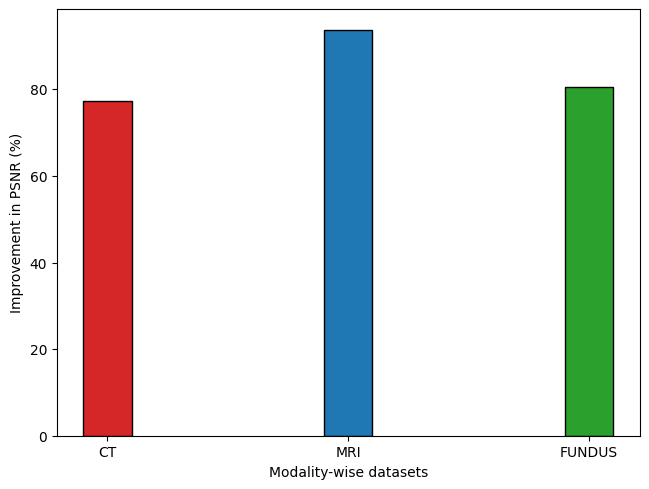}
    \caption{Percentage improvement in PSNR performance of DPL over the baseline U-Net across CT, MRI, and fundus datasets under Gaussian noise.}
    \label{fig:compare-dpl-baseline}
\end{figure}

The DPL architecture is highly flexible, allowing the use of either identical or distinct neural networks for its noise estimator, context estimator, and fusion reconstructor components. DPL demonstrates strong performance against additive noise such as Gaussian noise and remains competitive against multiplicative noise like speckle. Figure \ref{fig:mean-var-res} illustrates the mean and variance of the PSNR results for the baseline U-Net, RED-CNN, and DPL models across the CT, MRI, OCT, and Fundus datasets under Gaussian noise. Among the three models, RED-CNN exhibits the lowest variance, while DPL shows higher variability suggesting that its performance is more sensitive to image characteristics. Notably, DPL achieves its largest improvement over RED-CNN on the Fundus dataset. Figure \ref{fig:compare-dpl-baseline} presents the percentage improvement in PSNR when using DPL compared to the baseline U-Net across the CT, MRI, and Fundus datasets under Gaussian noise. The results show that DPL outperforms the baseline on approximately 80\% of the images, indicating consistent and robust improvements rather than isolated outlier gains that artificially inflate the mean PSNR.

Notably, the DPL model demonstrates the capability to be trained across all imaging modalities simultaneously-an important advancement, as most existing research is constrained to single-modality approaches. This flexibility may reduce the need for developing separate models for each modality in clinical settings. However, the current study has a few limitations. All models were trained for only 10,000 epochs due to computational constraints; extended training may further improve performance. Additionally, while DPL is more computationally demanding than single-network architectures, it offers the significant advantage of handling multiple modalities and various noise types within a unified framework. Another limitation lies in the relatively small OCT dataset, where DPL did not yield promising results. Future work may explore the use of larger OCT datasets to enable more reliable benchmarking. A comparison of training times across models is presented in Table \ref{tab:computational-cost}.

\begin{table}[h!]
    \centering
    \caption{Comparison of total training times for baseline U-Net, RED-CNN, and DPL models trained on 500 CT samples with Gaussian noise.}
    \begin{tabular}{c c}
        \hline
        Algorithm & Total Training time (hrs) \\
        \hline
        Baseline(UNet) & 16 \\
        RED-CNN & 17  \\
        DPL (Ours) & 46.66 \\
        \hline
    \end{tabular}
    \label{tab:computational-cost}
\end{table}

\section{Conclusion} \label{conclusion}
In this study, we introduced the Dual-Path Learning model, a flexible and adaptable architecture capable of leveraging different network designs based on specific requirements to effectively denoise medical images. DPL employs dual-pathway learning, simultaneously extracting information from both noise characteristics and contextual features within the image. The model was extensively evaluated across multiple noise types and imaging modalities, demonstrating strong robustness and generalizability. This comprehensive evaluation underscores the importance of developing denoising methods that are versatile and effective across diverse conditions, rather than being limited to single modality or noise-specific solutions. Future work will focus on improving the noise estimator for speckle and additive white Gaussian noise (AWGN) and expanding evaluations using larger datasets with additional imaging modalities.

\printbibliography
\end{document}